\documentclass[showpacs,twocolumn,preprintnumbers,showpacs,showkeys,superscriptaddress,amsmath,amssymb,nofootinbib]{revtex4}
\usepackage{amsmath}
\usepackage{latexsym}
\usepackage{color}
\usepackage{latexsym}
\usepackage{amsmath}
\usepackage{amssymb}
\usepackage{eufrak}
\usepackage{euscript}
\usepackage{pstricks}
\usepackage{graphics}
\usepackage{graphicx}
\usepackage{picture}

\newcommand{\be}{\begin{equation}}
\newcommand{\ee}{\end{equation}}
\newcommand{\ba}{\begin{eqnarray}}
\newcommand{\ea}{\end{eqnarray}}

\def\ni{\noindent}

\def\no{\nonumber}
\def\uma{\rm 1\!\!\hskip 1 pt l}
\begin{document}

\title{\Large{Spontaneous symmetry breaking and masses numerical results \\ in DFR noncommutative space-time }}

\author{M. J. Neves}\email{mariojr@ufrrj.br}
\affiliation{Grupo de F\'{i}sica Te\'orica e Matem\'atica F\'{i}sica, Departamento de F\'{\i}sica, Universidade Federal Rural do Rio de Janeiro, BR 465-07, 23890-971, Serop\'edica, RJ, Brazil}
\author{Everton M. C. Abreu}\email{evertonabreu@ufrrj.br}
\affiliation{Grupo de F\'{i}sica Te\'orica e Matem\'atica F\'{i}sica, Departamento de F\'{\i}sica, Universidade Federal Rural do Rio de Janeiro, BR 465-07, 23890-971, Serop\'edica, RJ, Brazil}
\affiliation{Departamento de F\'{i}sica, Universidade Federal de Juiz de Fora, 36036-330, Juiz de Fora, MG, Brazil}

\date{\today}


\begin{abstract}
\ni With the elements of the Doplicher, Fredenhagen and Roberts (DFR) noncommutative formalism, we have constructed the standard electroweak model.
We have introduced the spontaneous symmetry breaking and the hypercharge in DFR framework.
The electroweak  symmetry breaking was analyzed and the masses of the new bosons were computed.
\end{abstract}

\pacs{11.15.-q; 11.10.Ef; 11.10.Nx}

\keywords{DFR noncommutativity, electroweak standard model}

\maketitle




The fact that we have infinities that destroy the final results of several calculations in QFT have motivated theoretical physicists to ask if a continuum space-time would be really necessary. One of the possible solutions would be to create a discrete space-time with a noncommutative (NC) algebra,
where the position coordinates would be promoted to operators $\hat{X}^{\mu}\,(\mu=0,1,2,3)$ and they must satisfy commutation relations
\begin{eqnarray}\label{xmuxnu}
[\,\hat{X}^\mu\,,\,\hat{X}^\nu\,]\,=\,i\,\ell \, \theta^{\mu\nu} \, \hat{\openone} \; ,
\end{eqnarray}
where $\ell$ is a length parameter, $\theta^{\mu\nu}$ is an antisymmetric constant matrix and $\hat{\openone}$ is the identity operator.  In this way, we would have a kind of fuzzy space-time where, from this commutator, we have an uncertainty in the position coordinate.
In order to put these ideas together, Snyder \cite{snyder47} have published the first work that considers the space-time as being NC.
However, the frustrated result have doomed Snyder's NC theory to years of ostracism \cite{yang47}. After the relevant result that the algebra obtained from string theory embedded in a magnetic background is NC, a new flame concerning noncommutativity (NCY) was rekindle \cite{seibergwitten99}.
One of the paths (the most famous at least) of introducing NCY is through the Moyal-Weyl product where the NC parameter,
{\it i.e.} $\theta^{\mu\nu}$, is an antisymmetric constant matrix, namely,
\begin{eqnarray}\label{ProductMoyal}
\left. f(x) \star g(x) =
e^{\frac{i}{2}\theta^{\mu\nu}\partial_{\mu}\partial^{\prime}_{\nu}}
f(x) \, \, g(x^{\prime}) \right|_{x^{\prime}=x} \; .
\end{eqnarray}
\ni However, at superior orders of calculations, the Moyal-Weyl product turns out to be
highly nonlocal. This fact forced us to work with low orders in $\theta^{\mu\nu}$. Although it keeps the translational invariance,
the Lorentz symmetry is not preserved \cite{Szabo03}. For instance, concerning the case of the hydrogen atom, it breaks the rotational symmetry of the model, which removes the degeneracy of the energy levels \cite{Chaichian}.  Other subjects where the objective is to introduce NC effects in gravity \cite{grav}, in anyon models \cite{anyons} and symmetries \cite{sym}.  NCY through path integrals and coherent states was devised in \cite{path}.  For more generalized NC issues and reviews, the interested reader can look at \cite{reviews} and the references therein.

One way  to work with the NCY was introduced by Doplicher, Fredenhagen and Roberts (DFR). They have considered the parameter $\theta^{\mu\nu}$ as an ordinary coordinate of the system \cite{DFR1,DFR2}. This extended and new NC space-time has ten dimensions: four relative to the Minkowski space-time and six relative to $\theta$-space. Recently, in \cite{Morita,alexei,jhep} it was demonstrated that the DFR formalism have in fact a canonical momentum associated with $\theta^{\mu\nu}$
\cite{alexei,saxell,Amorim1} (for a review, \cite{amo}). The DFR framework is characterized by a field theory constructed in a space-time with extra-dimensions $(4+6)$,
and which does not need necessarily the presence of a length scale $\ell$ localized into the six dimensions
of the $\theta$-space where, from (\ref{xmuxnu}), we can see that $\theta^{\mu\nu}$ has dimension of length-square,
when we make $\ell=1$. By taking the limit with no such scale, the usual algebra of
the commutative space-time is recovered. Besides the Lorentz invariance was recovered, we obviously hope that causality aspects in
QFT in this $\left(x+\theta\right)$ space-time must be preserved too \cite{AbNeves}.

\section{The NC Yang-Mills symmetry revisited}

Considering a NC Yang-Mills model in the DFR framework \cite{YMAbreuNeves2015}, we have analyzed  the
gauge invariance of the fermion action under the star gauge symmetry transformations. The fermion Lagrangian coupled to
NC gauge fields is given by
\begin{eqnarray}\label{LDiracDcov}
{\cal L}_{Spinor}=\bar{\psi} \star \Big(\,i\gamma^{\mu}D_{\mu}\star
+\,{{i}\over2}\,\Gamma^{\mu\nu}D_{\mu\nu}\star
-m \Big) \psi  \; ,
\end{eqnarray}
where $\Gamma^{\mu\nu}:=i\left[\gamma^{\mu},\gamma^{\nu}\right]/4$ is the rotation generator of the fermions,
directly attached to the $\theta$-space.

Here we have defined the NC covariant derivative as being $D_{\mu}\star=\partial_{\mu}+igA_{\mu}\star$,
and $D_{\mu\nu}\star$ is a new antisymmetric star-covariant
derivative associated to the s $\theta$-space
\begin{eqnarray}\label{Dmunu}
D_{\mu\nu}\star:=\lambda\partial_{\mu\nu}+i g^{\prime}B_{\mu\nu}\star \; ,
\end{eqnarray}
where the field $B_{\mu\nu}$ is an antisymmetric tensor $(B_{\mu\nu}=-B_{\nu\mu})$ with six independents components.
The Lagrangian (\ref{LDiracDcov}) is manifestly invariant under star-gauge transformations
\begin{eqnarray}\label{PsiABgaugetransf}
\psi \hspace{0.1cm} \longmapsto \hspace{0.1cm}
\psi^{\prime} \! &=& \! U \star \psi \; ,
\nonumber \\
A_{\mu} \hspace{0.1cm} \longmapsto \hspace{0.1cm}
A_{\mu}^{\prime} \! &=& \! U \star A_{\mu}\star U^{\dagger}
-\frac{i}{g}\left(\partial_{\mu}U\right) \star U^{\dagger} \; ,
\nonumber \\
B_{\mu\nu} \hspace{0.1cm} \longmapsto \hspace{0.1cm}  B_{\mu\nu}^{\prime} \! &=& \! U \star B_{\mu\nu} \star U^{\dagger}-\frac{i}{g^{\prime}}\left(\lambda\partial_{\mu\nu}U\right) \star U^{\dagger} \; ,
\hspace{0.6cm}
\end{eqnarray}
since we have imposed that the element $U$ is an unitary-star, that is, $U^{\dagger}\star U=\uma$.
The Moyal product of two unitary matrix fields is always unitary, but in general
$\mbox{det}(U \, \star \, U^{\dagger}) \neq \mbox{det}(U) \, \star \, \mbox{det}(U^{\dagger})$, that is, $\det U \neq 1$.
Therefore, the group that represent the previous star gauge is unitary but not special, say $U^{\star}(N)$. The structure
of $U^{\star}(N)$ is the composition $U^{\star}(N)=U_{N}^{\star}(1) \times SU^{\star}(N)$ of a NC Abelian group with another
NC special unitary group. In gauge symmetry (\ref{PsiABgaugetransf}), we have obtained two NC gauge sectors: the first with a vector gauge
field, and the second one with tensor gauge field.   Hence, this gauge symmetry is composite of two unitary groups, say
$U^{\star}(N)_{A^{\mu}} \times U^{\star}(N)_{B^{\mu\nu}}$, and $U$ is the element of both groups.
The gauge fields $\left(A_{\mu},B_{\mu\nu}\right)$ are hermitian and it can be expanded in terms of the Lie algebra
generators in the adjoint representation as $A_{\mu}=A_{\mu}^{0} \, {\uma_{N}} +A_{\mu}^{a}t^{a}$ and
$B_{\mu\nu}=B_{\mu\nu}^{0} \, {\uma_{N}} +B_{\mu\nu}^{a}t^{a}$ where, by satisfying the Lie algebra commutation
relation we have that $\left[t^{a},t^{b}\right]=if^{abc}t^{c}$ $(a,b,c=1,\cdots, N^2-1)$. The fields $A_{\mu}^{0}$ and $B_{\mu\nu}^{0}$
come from the Abelian part of the group $U^{\star}(N)$, while the components $A_{\mu}^{a}$ and $B_{\mu\nu}^{a}$ are attached
to non-Abelian part of $U^{\star}(N)$. The fermion field $\psi$ is the column matrix of components
$\psi_{i}=\left(\psi_{1},\psi_{2}, \cdot \cdot \cdot, \psi_{N}\right)$ that lives in the fundamental
representation of the Lie algebra.

The dynamics in the gauge sector is introduced by the star commutators
\begin{eqnarray}\label{DmuDnuFmunu}
F_{\mu\nu} \!\! &=& \!\! -\frac{i}{g} \left[D_{\mu},D_{\nu}\right]_{\star}
=\partial_{\mu}A_{\nu}-\partial_{\nu}A_{\mu}+i g \, \left[A_{\mu},A_{\nu}\right]_{\star}  \; ,
\nonumber \\
G_{\mu\nu\rho\sigma} \!\! &=& \!\! -\frac{i}{g^{\prime}} \left[D_{\mu\nu} , D_{\sigma\rho} \right]_{\star}
=\lambda\partial_{\mu\nu}B_{\rho\sigma}
-\lambda\partial_{\rho\sigma}B_{\mu\nu}
\nonumber \\
&&+ig^{\prime}\left[B_{\mu\nu}, B_{\rho\sigma}\right]_{\star} \; .
\end{eqnarray}
By construction, they have the gauge transformations
\begin{eqnarray}\label{transfgaugeFmunu}
F_{\mu\nu} \; \longmapsto \; F_{\mu\nu}^{\;\prime} \!\! &=& \!\! U \star F_{\mu\nu} \star U^{\dagger} \; ,
\nonumber \\
G_{\mu\nu\rho\sigma} \; \longmapsto \;
G^{\;\prime}_{\mu\nu\rho\sigma} \!\! &=& \!\! U \star G_{\mu\nu\rho\sigma} \star U^{\dagger} \; .
\end{eqnarray}
%
%
Therefore, we have a Lagrangian for the gauge fields given by
\begin{eqnarray}\label{LGauge}
{\cal L}_{Gauge}&=&
-\frac{1}{4}\,\mbox{tr}_{N}\left(F_{\mu\nu}\star F^{\mu\nu}\right)
-\frac{1}{4}\,\mbox{tr}_{N}\left(G_{\mu\nu\rho\sigma}\star
G^{\mu\nu\rho\sigma}\right)
\nonumber \\
&&-\frac{1}{2}\,\mbox{tr}_{N}\left(F_{\mu\nu} \star G^{\mu\rho\nu}_{\hspace{0.5cm}\rho}
\right) \; .
\end{eqnarray}

\section{The Model $U_{L}^{\star}(2) \times U_{R}^{\star}(1) \times U_{L}(2)_{B^{\mu\nu}} \times U_{R}(1)_{X^{\mu\nu}}$}

Based on the symmetry $U^{\star}(N)_{A^{\mu}} \times U^{\star}(N)_{B^{\mu\nu}}$ we will tconstruct an electroweak model in the
NC DFR framework. The candidate for this goal is the composite group
$U_{L}^{\star}(2)_{A^{\mu}} \times U_{R}^{\star}(1)_{B^{\mu}} \times U_{L}(2)_{B^{\mu\nu}} \times U_{R}(1)_{X^{\mu\nu}}$, in which we have sectors
left and right-handed concerning  gauge vector fields, and the analogous one for the gauge tensor fields. This composite model would be version of the
Glashow-Salam-Weinberg model for electroweak interaction
in the context of DFR NCY.
%
%
Firstly, we will define the fermions doublets, neutrinos and leptons left-handed, that transforms in the fundamental
representation of $U_{L}^{\star}(2)$, and in the anti-fundamental representation of $U_{R}^{\star}(1)$ as
\begin{eqnarray}
\Psi_{L}=
\left(
  \begin{array}{c}
    \nu_{\ell L} \\
    \ell_{L} \\
  \end{array}
\right)
\longmapsto \Psi_{L}^{\prime}= U
\star \Psi_{L} \star V_{2}^{-1} \; ,
\end{eqnarray}
where $U$ is the element of any group $U_{L}^{\star}(2)$, and $V_{2}$ is the element of $U_{R}^{\star}(1)$.
For the right sector $U_{R}^{\star}(1)$, the fermions transformation  in the anti-fundamental representation as
\begin{eqnarray}
\ell_{R} \longmapsto \ell_{R}^{\prime}= \ell_{R} \star V_{2}^{-1} \; .
\end{eqnarray}
%
The covariant derivatives acting on fermions in the left and right-sectors of the model are defined by
\begin{eqnarray}
D_{L\mu}\Psi_{L}\!\! &=& \!\! \partial_{\mu}\Psi_{L}+ig_{1}A_{\mu} \star \Psi_{L}-i J_{L} g_{1}^{\prime}\, \Psi_{L} \star B_{\mu} \; ,
\nonumber \\
D_{L\mu\nu}\Psi_{L}\!\! &=& \!\! \lambda \, \partial_{\mu\nu} \Psi_{L} + ig_{2}B_{\mu\nu} \star \Psi_{L}
- i J_{L} g_{2}^{\prime} \, \Psi_{L} \star X_{\mu\nu} \; ,
\nonumber \\
D_{R\mu}\ell_{R} \!\! &=& \!\! \partial_{\mu}\ell_{R} - i J_{R} g_{1}^{\prime} \, \ell_{R} \star B_{\mu} \; ,
\nonumber \\
D_{R\mu\nu}\ell_{R} \!\! &=& \!\! \lambda \, \partial_{\mu\nu}\ell_{R} - i J_{R} g_{2}^{\prime} \, \ell_{R} \star X_{\mu\nu} \; ,
\end{eqnarray}
where $A_{\mu}\!=\!A_{\mu}^{\,0}\,{\uma_{2}}+A_{\mu}^{a} \, \frac{\sigma^{a}}{2}$ and $B_{\mu\nu}\!=\!B_{\mu\nu}^{\,0}\,{\uma_{2}}
+B_{\mu\nu}^{a} \, \frac{\sigma^{a}}{2}$ are the non-Abelian gauge fields of $U_{L}^{\star}(2)_{A^{\mu}}$ and $U_{L}^{\star}(2)_{B^{\mu\nu}}$,
and $B_{\mu}$ and $X_{\mu\nu}$ are the Abelian
gauge fields of $U_{R}^{\star}(1)_{B^{\mu}}$ and $U_{R}^{\star}(1)_{X^{\mu\nu}}$. We have used the symbol $J$ as the generator of $U_{R}^{\star}(1)$.
Imposing the gauge transformations analogously  to (\ref{PsiABgaugetransf}), we can construct the leptons Lagrangian invariant under previous gauge
transformations
%
%
%
%
%
\begin{equation}\label{Lleptons}
{\cal L}_{Leptons}=\bar{\Psi}_{L}\star i\gamma^{\mu}D_{L\mu} \star \Psi_{L}
+\bar{\ell}_{R}\star i\gamma^{\mu}D_{R\mu} \star \ell_{R} \; .
\end{equation}
%
%
%
The introduction of left and right handed components vanishes the terms of propagation in the $\theta$-space,
and all interactions of the fermions with the sector of the gauge tensor fields. This puzzle can be bypassed when
we introduce Yukawa interactions in the Higgs sector to break the gauge symmetry of $U_{L}^{\star}(2)_{B^{\mu\nu}}$.

In the sector gauge fields, the field strength tensors of the bosons are defined by
\begin{eqnarray}\label{FieldStrength}
F_{\mu\nu} \!\! &=& \!\! \partial_{\mu}A_{\nu}-\partial_{\nu}A_{\mu}+ig_{1} \left[A_{\mu}, A_{\nu} \right]_{\star}  \; ,
\nonumber \\
H_{\mu\nu} \!\! &=& \!\! \partial_{\mu}B_{\nu}-\partial_{\nu}B_{\mu}+i J\, g_{1}^{\prime} \, \left[B_{\mu} , B_{\nu}\right]_{\star} \; ,
\nonumber \\
G_{\mu\nu\rho\lambda} \!\! &=& \!\! \lambda\partial_{\mu\nu}B_{\rho\lambda}-\lambda\partial_{\rho\lambda}B_{\mu\nu}
+ ig_{2} \left[B_{\mu\nu}, B_{\rho\lambda} \right]_{\star} \; ,
\nonumber \\
X_{\mu\nu\rho\lambda} \!\! &=& \!\! \lambda\partial_{\mu\nu}X_{\rho\lambda}-\lambda\partial_{\rho\lambda}X_{\mu\nu}+i J \, g_{2}^{\prime} \left[X_{\mu\nu} , X_{\rho\lambda}\right]_{\star} .
\hspace{0.5cm}
\end{eqnarray}
%
%
%
The Lagrangian of the invariant gauge fields invariant is given by
\begin{eqnarray}\label{Lgauge}
{\cal L}_{Gauge}\!\!\!&&=-\frac{1}{2} \,\mbox{tr}\left(F_{\mu\nu}\star F^{\mu\nu}\right)
-\frac{1}{4} \, H_{\mu\nu} \star H^{\mu\nu}
\nonumber \\
&&-\frac{1}{2} \, \mbox{tr}\left(G_{\mu\nu\rho\lambda} \star G^{\mu\nu\rho\lambda}\right)
-\frac{1}{4} \, X_{\mu\nu\rho\lambda} \star X^{\mu\nu\rho\lambda}
\nonumber \\
&&
- \, \mbox{tr}\left(F_{\mu\nu} \star G^{\mu\rho\nu\hspace{0.15cm}}_{\hspace{0.45cm}\rho}\right)
-\frac{1}{2} \, H_{\mu\nu} \star X^{\mu\rho\nu}_{\hspace{0.48cm}\rho} \; .
\end{eqnarray}
%
%
%

The interaction terms that emerge in (\ref{Lleptons}) reveal that the leptons
and neutrinos can interact with the gauge fields components. Initially, we will write these interactions as
\begin{eqnarray}\label{LintAY}
&&{\cal L}_{Leptons-Gauge}^{\, int}
=-\bar{\Psi}_{L} \star \gamma^{\mu}\left( \, g_{1}A_{\mu}^{3}I^{3}+g_{1} \, A_{\mu}^{0}
\right. \nonumber \\
&& \left. - J_{L} \, g_{1}^{\prime} \, B_{\mu} \,\right) \star \Psi_{L}
+\bar{\ell}_{R}\star\gamma^{\mu}\left(J_{R} \, g_{1}^{\prime} B_{\mu} \right)\star \ell_{R} + \cdots
\; , \hspace{0.8cm}
\end{eqnarray}
where we have defined $I^{3}=\sigma^{3}/2$, for simplicity.
Looking these terms, we can ask who is hypercharge generator
of the model. In the next section, we will use a Higgs mechanism to define the hypercharge in DFR space-time.

\section{The first SSB and the hypercharge}

To identify the hypercharge, we will introduce the first Higgs sector
coupled to the NC Abelian gauge vector field, and consequently, we will eliminate the residual symmetry $U^{\star}(1)$,
i.e., the Abelian subgroup of $U_{L}^{\star}(2)_{A^{\mu}}$. This Higgs sector is also coupled to the gauge tensor field
of the non-Abelian sector $U_{L}^{\star}(2)_{B^{\mu\nu}}$ to give for the new anti-symmetrical bosons. We will denote
this Higgs field as the Higgs-one $\Phi_{1}$. After this first spontaneous symmetry breaking (SSB), we will obtain
\\
$U_{L}^{\star}(2) \times U_{R}^{\star}(1) \times U_{L}^{\star}(2)_{B^{\mu\nu}} \times U_{R}^{\star}(1)_{X^{\mu\nu}}
\stackrel{\langle \Phi_{1} \rangle}{\longmapsto} SU_{L}^{\star}(2) \times U_{Y}^{\star}(1) \times U^{\star}(1) \times U_{R}^{\star}(1)$.
\\
To do that, we will introduce the Higgs-one Lagrangian
\begin{eqnarray}\label{LHiggs1}
{\cal L}_{Higgs}^{\, (1)} \!\! &=& \!\! \left(D_{\mu}\Phi_{1}\right)^{\dagger} \star D^{\mu} \Phi_{1}
+ \frac{1}{2} \, \left(D_{\mu\nu} \Phi_{1}\right)^{\dagger} \star D^{\mu\nu}\Phi_{1}
\no \\
&& \!\! -\mu_{1}^2 \left(\Phi_{1}^{\,\dagger} \star \Phi_{1}\right)-g_{H1} \left(\Phi_{1}^{\,\dagger} \star \Phi_{1}\right)^{2} \; ,
\end{eqnarray}
%
%
%
%
%
%
%
where $\mu_{1}$, $g_{H1}$ are real parameters.
The covariant derivatives of (\ref{LHiggs1}) act on the Higgs-one as
\begin{eqnarray}\label{DmuPhi1}
D_{\mu} \Phi_{1} \!\! &=& \!\! \partial_{\mu}\Phi_{1}+ig_{1}A_{\mu}^{\; 0} \star \Phi_{1}-i J g_{1}^{\prime}\, \Phi_{1} \star B_{\mu} \; ,
\nonumber \\
D_{\mu\nu} \Phi_{1} \!\! &=& \!\! \lambda\partial_{\mu\nu}\Phi_{1}+ig_{2} B_{\mu\nu}^{\; 0} \star \Phi_{1}
+ig_{2}B_{\mu\nu}^{a}\frac{\sigma^a}{2}\star\Phi_{1} \; .
\hspace{0.5cm}
\end{eqnarray}
The field $\Phi_{1}$ is a scalar doublet of both groups $U_{L}^{\star}(2)$.
In the antisymmetric sector, the Higgs-one transforms into the fundamental representation of $U_{L}^{\star}(2)_{B^{\mu\nu}}$
as
\begin{eqnarray}\label{transfPhi1U}
\Phi_{1}=
\left(
\begin{array}{c}
\phi_{1}^{ \, (+)} \\
\phi_{1}^{\, (0)} \\
\end{array}
\right)
\longmapsto \Phi_{1}^{\prime}=U \star \Phi_{1} \; .
\end{eqnarray}
In the NC Abelian sector, $\Phi_{1}$ transforms in the
fundamental representation of $U^{\star}(1)$, and in the anti-fundamental representation of
$U_{R}(1)$ as
\begin{equation}\label{transfPhi1VV}
\Phi_{1} \longmapsto \Phi_{1}^{\; \prime}= V_{1} \star \Phi_{1} \star V_{2}^{-1} \; ,
\end{equation}
where $V_{1}$ is the element of the subgroup $U^{\star}(1)$.


When the Higgs potential acquires a non-trivial vacuum expected value (VEV), say $v_1 \neq 0$, we can choose the usual parametrization
in the unitary gauge,
%
%
so the massive terms in (\ref{LHiggs1}) are given by
\begin{eqnarray}\label{LHiggs1Masses}
{\cal L}_{Mass}^{(1)} \!\! &=& \!\! \frac{1}{2}\, m_{B^{\pm}}^{2} \, B_{\mu\nu}^{+}B^{\mu\nu -}
+\frac{v_{1}^{2}}{2} \left(\phantom{\frac{1}{2}} \!\!\!\!\! g_{1} A_{\mu}^{0}- J g_{1}^{\prime} B_{\mu}\right)^{2}
\nonumber \\
&&\!\!+\frac{v_{1}^{2}}{4} \left( -\frac{1}{2} \,  g_{2} B_{\mu\nu}^{3} + g_{2} B_{\mu\nu}^{0} \right)^{2} \; .
\end{eqnarray}
Notice that the emergence of a new charged field $\sqrt{2} \, B_{\mu\nu}^{\pm}=B_{\mu\nu}^{1}\mp iB_{\mu\nu}^{2}$,
where the mass is $m_{B^{\pm}}=g_{2}v_{1}/2$. To define the hypercharge, the other mass terms suggest us to
introduce the orthogonal transformations
\begin{eqnarray}\label{transfA0CGY}
A_{\mu}^{0} \!\! &=& \!\! \cos\alpha \, G_{\mu}+ \sin\alpha \, Y_{\mu} \; ,
\nonumber \\
B_{\mu}\!\! &=& \!\! -\sin\alpha \, G_{\mu}+\cos\alpha \, Y_{\mu} \; ,
\nonumber \\
B_{\mu\nu}^{0} \!\! &=& \!\! \cos\beta \, G_{\mu\nu}+ \sin\beta \, Y_{\mu\nu} \; ,
\nonumber \\
X_{\mu\nu}\!\! &=& \!\! -\sin\beta \, G_{\mu\nu}+\cos\beta \, Y_{\mu\nu} \; ,
\end{eqnarray}
where $\alpha,\beta$ are the mixing angles, and $\tan\alpha=J g_{1}^{\prime}/g_{1}$.
Here, the fields $Y$ set the gauge fields associated to the hypercharge generator, where
we define $g^{\prime} \, Y_{\Phi_{1}}= g_{2} \sin\beta$, and the hypercharge of the Higgs is
$Y_{\Phi_{1}}=1/2$. Therefore, the Lagrangian (\ref{LHiggs1Masses}) is rewritten as
\begin{eqnarray}\label{LHiggs1MassesGmu}
{\cal L}_{Mass}^{(1)} \!\! &=& \!\! \frac{1}{2} \, m_{B^{\pm}}^{2} \, B_{\mu\nu}^{+}B^{\mu\nu -}
+\frac{1}{2} \, m_{G_{\mu}}^{2} G_{\mu}G^{\mu}
\nonumber \\
&&\hspace{-1cm}+\frac{v_{1}^{2}}{4} \left[g_{2}\cos\beta \, G_{\mu\nu}
+ \frac{1}{2} \left(g^{\prime} \, Y_{\mu\nu} - g_{2} \, B_{\mu\nu}^{3}\right) \right]^{2} ,
\end{eqnarray}
where we obtain the mass of $G_{\mu}$ given by the expression
\begin{eqnarray}\label{massesGmuGmunu}
m_{G_{\mu}}=v_{1} \, \sqrt{g_{1}^{2}+(Jg_{1}^{\prime})^{\, 2}}=\frac{g_{1} \, v_{1}}{\cos\alpha} \; .
\end{eqnarray}
The last term in (\ref{LHiggs1MassesGmu}) suggests us to make the second orthogonal transformation
\begin{eqnarray}\label{transfBYmunuZmunu}
B_{\mu\nu}^{3} \!\! &=& \!\! \cos\theta_{2} \, Z_{\mu\nu}+ \sin\theta_{2} \, A_{\mu\nu}
\nonumber \\
Y_{\mu\nu}\!\!&=& \!\!-\sin\theta_{2} \, Z_{\mu\nu}+\cos\theta_{2} \, A_{\mu\nu} \; ,
\end{eqnarray}
where $\theta_{2}$ is another mixing angle, satisfying the condition $\tan\theta_{2}=2\sin\beta$,
we obtain
\begin{eqnarray}\label{LHiggs1MassesZG}
{\cal L}_{Mass}^{(1)} \!\!&=& \!\! \frac{1}{2} \, m_{B^{\pm}}^{2} \, B_{\mu\nu}^{+}B^{\mu\nu -}
\!+\frac{1}{2} \, m_{G_{\mu}}^{2} G_{\mu}G^{\mu}
\nonumber \\
&& \hspace{-0.8cm}+\frac{g_{2}^{2}v_{1}^{2}}{4}\left(\cos\beta \, G_{\mu\nu}
- \frac{1}{2} \, \sec\theta_{2} \, Z_{\mu\nu} \right)^{2}  \, \; .
\end{eqnarray}
We will diagonalize the last term in order to obtain the mass of $Z^{\mu\nu}$ is
$m_{Z_{\mu\nu}}\!=\sqrt{5}\, g_{2} v_{1}/2$, while $G_{\mu\nu}$ remains massless in this SSB.
Comparing the masses of $B^{\pm}$ and $Z_{\mu\nu}$,
we will obtain the relation $m_{Z_{\mu\nu}}=\sqrt{5} \, m_{B^{\pm}}$.

%
%
%
%
%
%
The interactions between leptons-neutrinos and gauge vector bosons in (\ref{LintAY}) can be written in terms of the
fields $G^{\mu}$ and $Y^{\mu}$ to identify the hypercharge generators of the left-right sectors as
$J_{R} g_{1}^{\prime} \cos\alpha= -g \, Y_{R}$, $g_{1}\sin\alpha -J_{L} g_{1}^{\prime} \cos\alpha=+g \, Y_{L}$.
These definitions give us
\begin{eqnarray}\label{LintGY}
{\cal L}_{Leptons-Gauge}^{\, int} \!\! &=& \!\! -\bar{\Psi}_{L} \star \gamma^{\mu}\left( \, g_{1}A_{\mu}^{3} \, I^{3}
+ g \, Y_{L} Y_{\mu} \, \right) \star \Psi_{L}
\nonumber \\
&&\hspace{-0.5cm} +\bar{\ell}_{R}\star\gamma^{\mu} \left(- g \, Y_{R} \, Y_{\mu} \right) \star \ell_{R} + \cdots
\end{eqnarray}
Here we are ready to discover how the mixing $A_{\mu}^{3}-Y_{\mu}$ defines the physical particles $Z^{0}$ and massless photon,
and posteriorly, the electric charge of the particles. To this end, we need to break the resting symmetry of this SSB.
We will make this second mechanism in the next section.


%
\section{The electroweak symmetry breaking}

Until now we had constructed a model for NC electroweak interaction using a Higgs sector to eliminate the residual symmetry $U^{\star}(1)$,
and we have defined the hypercharge of the Abelian sector. Now we are going to introduce a second Higgs sector $\Phi_{2}$ in order to break the
electroweak symmetry. We write the Lagrangian of the second Higgs-$\Phi_{2}$ as the scalar sector, that is,
%
%
%
\begin{eqnarray}\label{LHiggs2}
{\cal L}_{Higgs}^{(2)} \!\! &=& \!\! \left(D_{\mu}\Phi_{2}\right)^{\dagger} \star D^{\mu} \Phi_{2}
+ \frac{1}{2} \, \left(D_{\mu\nu} \Phi_{2} \right)^{\dagger} \star D^{\mu\nu}\Phi_{2}
\no \\
&& \!\! -\mu_{2}^2 \left(\Phi_{2}^{\dagger} \star \Phi_{2}\right)-g_{H2} \left(\Phi_{2}^{\dagger} \star \Phi_{2}\right)^{2} \; ,
\end{eqnarray}
%
%
%
where $\mu_{2}$ and $g_{H2}$ are real parameters.
The field $\Phi_{2}$ is a complex scalar doublet that has the gauge transformation analogous to that
from $\Phi_{1}$. The covariant derivatives act on $\Phi_{2}$ as
\begin{eqnarray}
D_{\mu} \Phi_{2} \!\! &=& \!\! \partial_{\mu}\Phi_{2} + ig_{1} A_{\mu}^{0} \star \Phi_{2}
+ ig_{1} A_{\mu}^{\, a} \, \frac{\sigma^{a}}{2} \star \Phi_{2} \; ,
\nonumber \\
D_{\mu\nu} \Phi_{2} \!\! &=& \!\! \lambda \, \partial_{\mu\nu}\Phi_{2} + ig_{2}B_{\mu\nu}^{0} \star \Phi_{2}
-iJ g_{2}^{\prime} \Phi_{2} \star X_{\mu\nu} \; . \hspace{0.6cm}
\end{eqnarray}
Using the transformations (\ref{transfA0CGY}), the term in $Y_{\mu}$ in the covariant derivative suggests that
$g\, Y_{\Phi_{2}}:=g_{1} \, \sin\alpha$. We use a similar parametrization to first SSB to obtain the result
\begin{eqnarray}\label{massesBWZ}
{\cal L}_{Mass}^{\,(2)} \!\! &=& \!\! m_{W^{\pm}}^{2} \, W_{\mu}^{\;+}W^{\mu-}
+\frac{v^{2}}{4} \left( \!\!\!\!\! \phantom{\frac{1}{2}} \, g_{2} B_{\mu\nu}^{\, 0}- J g_{2}^{\prime} \, X_{\mu\nu} \right)^{2}
\nonumber \\
&&\!\! +\frac{v^{2}}{2} \left[g_{1}\cos\alpha \, G_{\mu}
+\frac{1}{2} \left( \phantom{\frac{1}{2}} \!\!\!\! g \, Y_{\mu}-g_{1} \, A_{\mu}^{3} \right) \right]^{2} \; ,
\hspace{0.8cm}
\end{eqnarray}
where $v$ is the VEV that defines the scale for this SSB.
As in the usual case, the mass of $W^{\pm}$ is $m_{W^{\pm}}=g_{1}v/2$.
%
%
The mass terms of the neutral bosons in (\ref{massesBWZ}) motivate us to introduce the orthogonal transformations
\begin{eqnarray}\label{transfAYZ}
A_{\mu}^{3} \! &=& \! \cos\theta_{W} \, Z_{\mu} + \sin\theta_{W} \, A_{\mu}
\nonumber \\
Y_{\mu} \! &=& \! -\sin\theta_{W} \, Z_{\mu} + \cos\theta_{W} \, A_{\mu} \; ,
\end{eqnarray}
%
%
where we have that $\tan\theta_{W}=g/g_{1}$, and $\tan\beta=J g_{2}^{\prime}/g_{2}$.   We can find all the massive terms in this Lagrangian
\begin{eqnarray}\label{Lmass}
&&{\cal L}_{Mass}= m_{W^{\pm}}^{2}\, W_{\mu}^{\;+}W^{\mu-}
+\frac{1}{2} \, m_{B^{\pm}}^{2} \, B_{\mu\nu}^{+}B^{\mu\nu -}\!
\nonumber \\
&&+\frac{g_{1}^{2}v^{2}}{2} \left( \frac{1}{2} \, \sec\theta_{W} Z_{\mu} - \cos\alpha \, G_{\mu} \right)^{2}
\!\!\!+\frac{1}{2} \, \frac{g_{1}^{2}v_{1}^{2}}{\cos^{2}\alpha}  \, G_{\mu}G^{\mu}
\nonumber \\
&&+\frac{1}{4} \, \frac{g_{2}^{2}v^{2}}{\cos^2\beta} \, G_{\mu\nu}G^{\mu\nu}\!+\frac{1}{4} \, m_{Z_{\mu\nu}}^{\, 2}  \, Z_{\mu\nu}Z^{\mu\nu} \; .
\end{eqnarray}
Here we have taken into account the mass terms from the first SSB of the Higgs-$\Phi_{1}$.
Notice that the fields $A_{\mu}$ and $A_{\mu\nu}$ are not present in the Lagrangian (\ref{massesBWZ}).
They are the massless gauge fields remaining in the model after SSBs, namely, we have the final symmetry
$U_{L}^{\star}(2)_{A^{\mu}} \times \times U_{R}^{\star}(1)_{B^{\mu}} \times U_{L}^{\star}(2)_{B^{\mu\nu}} \times U_{R}^{\star}(1)_{X^{\mu\nu}}  \stackrel{\langle \Phi_{1} \rangle}{\longmapsto} SU_{L}^{\star}(2) \times U_{Y}^{\star}(1) \times U^{\star}(1) \times U_{R}^{\star}(1) \stackrel{\langle \Phi_{2} \rangle}{\longmapsto} U_{em}^{\star}(1) \times U^{\star}(1)_{A^{\mu\nu}}$, where $A_{\mu}$ is the photon field and
$A_{\mu\nu}$ is its antisymmetrical correspondent in the $\theta$-space.
It is important to explain that the $Z$-field, which came from (\ref{transfAYZ}) is not the $Z^{0}$-particle of the standard electroweak model.
The $Z^{0}$-particle will be defined by means of the mixing with the $G$-field in (\ref{Lmass}).
Since we had established the scale $v_{1} \gg v$,
we have diagonalized the mixing term $Z-G$, so the masses of $Z$, $G$ and their antisymmetric pairs up to the second order in $v/v_{1}$ are given by
\begin{eqnarray}
m_{Z^{0}} \!\! &=& \!\! \frac{g_{1}v}{2\cos\theta_{W}} \left(1-\frac{v^{2}}{2 v_{1}^{2}} \cos^{4}\alpha + \ldots  \right)
\; ,
\nonumber \\
m_{G_{\mu}} \!\! &=& \!\! \frac{g_{1}v_{1}}{\cos\alpha}\left(1+\frac{v^{2}}{2 v_{1}^{2}} \cos^{4}\alpha +\ldots  \right) \; .
\end{eqnarray}
%
Substituting (\ref{transfAYZ}) into (\ref{LintGY}), we can identify the fundamental charge by the parametrization
\begin{eqnarray}\label{eg1g}
e=g_{1}\sin\theta_{W}=g\cos\theta_{W} \; ,
\end{eqnarray}
where the electric charge is given by $Q_{em}=I^{3}+Y$.
We will use the $VEV$ of this SSB as the electroweak scale, that is, $v\simeq 246 \, GeV$, and
considering the experimental value of $\sin^{2}\theta_{W} \simeq 0.23$, the masses of $W^{\pm}$ and $Z^{0}$
are estimated to give the values
\begin{eqnarray}\label{massesWZ}
m_{W^{\pm}} \!\! &=& \!\! \frac{37 \, \, GeV}{|\sin\theta_{W}|} \simeq 77 \, GeV
\quad , \quad
\nonumber \\
m_{Z^{0}} \!\! &=& \!\! \frac{74 \, \, GeV}{|\sin2\theta_{W}|} \left(1-\frac{v^{2}}{2 v_{1}^{2}} \cos^{4}\alpha + \ldots  \right)
\nonumber \\
&\simeq& \!\! 89 \, GeV \, \left(1-\frac{v^{2}}{2 v_{1}^{2}} \cos^{4}\alpha + \ldots  \right) \; .
\end{eqnarray}
To estimate the values for the masses of $Z_{\mu\nu}$, $B^{\pm}$ and $G_{\mu\nu}$,
we have to examine the $3$-line and $4$-line vertex of the bosons $B^{\pm}$ interacting with the $A^{\mu}$-photon.
Using the universality of the electromagnetic interaction, the coupling constant of these vertex is given by the fundamental charge, so we find the relation
$g_{2}=g_{1}=e\, \csc \theta_{W}$, and the $\alpha$-angle is connected to $\theta_{W}$ by $\sin\alpha=\tan\theta_{W}$, so we obtain
$\sin^{2}\alpha \simeq 0.33$. This is the result of the NC standard model in the framework of $\theta^{\mu\nu}$-constant \cite{Chaichian2003}.
In the NC model, the scale of NCY has a lower bound of $\Lambda_{NC} \gtrsim 10^{3} \, GeV$, so we use this scale to represent the first $VEV$, that is, $v_{1} \simeq 1$ TeV. Consequently, the masses of the bosons $B^{\pm}$ and $Z_{\mu\nu}$ can be computed as
\begin{eqnarray}
m_{G_{\mu}} \simeq 770 \, GeV
\hspace{0.05cm} , \hspace{0.05cm}
m_{B^{\pm}} \!\! &\simeq& \!\! 310 \, GeV
\hspace{0.05cm} , \hspace{0.05cm}
m_{Z_{\mu\nu}} \simeq 699 \, GeV \; ,
\nonumber \\
\hspace{-1cm} m_{G_{\mu\nu}} \!\! &\simeq& \!\! \frac{154}{\cos\beta} \, GeV > 154 \, GeV \; .
\end{eqnarray}
%

%
%


In this way, we have analyzed some elements of the NC standard model such as the electroweak standard model.
Since the position and $\theta$ coordinates are independent variables, the Weyl-Moyal product keeps its associative property and it is the basic product, as usual in canonical NC models.

Hence, we have introduced new ideas and concepts in DFR formalism and we began with the construction of the symmetry group $U^{*}_L (2) \times U^{*}_R (1)$, which is the DFR version of the GSW model concerning the electroweak interaction, in order to introduce left and right-handed fermionic sectors.  Some elements such as covariant derivatives, gauge transformations and gauge invariant Lagrangians were constructed, and the interactions between leptons and gauge fields were discussed.

After that we have introduced the first Higgs sector to break one of the two Abelian NC symmetries in order to destroy the residual model's $U^{*}(1)$ symmetry. The spontaneous symmetry breaking was discussed and, in this way, the Higgs Lagrangian was introduced. We have seen that in the context of the NC DFR framework, the Abelian gauge field associated with $U^{*}(1)$ have acquired a mass term. Besides, thanks to the NC scenario, some fields are massive and others, massless. Also in the NC context we have obtained 3-line and 4-line vertex interactions and the renormalizability of the model was preserved.   
The residual symmetry $U^* (1)$ was eliminated via the use of the Higgs sector.

Moreover, we have introduced a second Higgs sector in order to break the electroweak symmetry and the masses of the old and new bosons were computed with the NC contributions. Since the Weinberg angle was identified as the basic angle to calculate the masses of the $W^{\pm}$ and $Z^0$, we have used the experimental value of the sine of the Weinberg angle in order to calculate the $W^{\pm}$ and $Z^0$ masses in an NC scenario. We have used the lower bound for the first SSB scale given by $v_{1} \simeq 1$ TeV. Finally, we have obtained the masses for new antisymmetric bosons of the DFR framework.


\bigskip


\ni E.M.C.A. thanks CNPq (Conselho Nacional de Desenvolvimento Cient\' ifico e Tecnol\'ogico), Brazilian scientific support federal agency, for partial financial support through Grants No. 301030/2012-0 and No. 442369/2014-0 and the hospitality of Theoretical Physics Department at Federal University of Rio de Janeiro (UFRJ), where part of this work was carried out.


\end{document}